\documentclass[12pt,nofootinbib,preprint,aps]{revtex4}

\usepackage{amssymb,amsmath,bm,bbm}
\usepackage{epsf}
\usepackage{epsfig}
\usepackage{afterpage}
\usepackage{longtable}
\usepackage{latexsym, mathrsfs}
\usepackage{color}

\newcommand{\cleqn}{\setcounter{equation}{0}}

\newcommand{\beq}{\begin{equation}}
\newcommand{\eeq}{\end{equation}}

\newcommand{\ODR}{{\overline{D}_{\rm R}}}

\newcommand{\DL}{{D_{\rm L}}}
\newcommand{\UL}{{U_{\rm L}}}

\def\bea{\begin{eqnarray}}
\def\eea{\end{eqnarray}}
\def\bcen{\begin{center}}
\def\ecen{\end{center}}
\def\btotn{B \to \tau \nu}
\def\btodtn{B \to D \tau \nu}

\def  \sigbr      {(\sigma \times BR) }

\def  \et  {\not\!\! E_T} 

\begin{document}

%\renewcommand{\thefootnote}{\fnsymbol{footnote}}

%%%%%  Title =========================%%
\title{Universality test of the charged Higgs boson
couplings at the LHC and at $B$-factories } 
%%%%%% Authors =======================%%
\author{Alan S. Cornell\footnote{alan.cornell@wits.ac.za}}
   \affiliation{National Institute for Theoretical Physics; School of
     Physics, University of the Witwatersrand, Wits 2050, South
     Africa} 
\author{Aldo Deandrea \footnote{deandrea@ipnl.in2p3.fr}}
   \affiliation{Universit\'e de Lyon, F-69622 Lyon, France; Universit\'e Lyon 1,\\
   CNRS/IN2P3, UMR5822 IPNL, F-69622 Villeurbanne Cedex, France, }
\author{Naveen Gaur \footnote{naveen@physics.du.ac.in}}
   \affiliation{Department of Physics \& Astrophysics, University of Delhi, Delhi -
    110007, India and \\
    Physics Division, National Center for Theoretical Sciences,
    Hsinchu, Taiwan}
\author{Hideo Itoh \footnote{hideo@post.kek.jp}}
    \affiliation{Institute for Cosmic Ray Research (ICRR), 
      University of Tokyo, 5-1-5, Kashiwanoha,
     Kashiwa city, Chiba, Japan and \\
     Theory Center, Institute of Particle and Nuclear Studies, KEK,
     1-1 Oho, Tsukuba, Ibaraki 305-0801, Japan}
\author{Michael Klasen \footnote{klasen@lpsc.in2p3.fr}}
    \affiliation{Laboratoire de Physique Subatomique et de Cosmologie,
     Universit\'e Joseph Fourier/CNRS-IN2P3/INPG, 53 Avenue des Martyrs,
     38026 Grenoble, France}
\author{Yasuhiro Okada \footnote{yasuhiro.okada@post.kek.jp}}
    \affiliation{
     Theory Center, Institute of Particle and Nuclear Studies, KEK,
     1-1 Oho, Tsukuba, Ibaraki 305-0801, Japan,
     and \\
     Department of Particle and Nuclear Physics, Graduate
     University of Advanced Studies (SOKENDAI) Tsukuba, Ibaraki 305-0801,
     Japan.}
%+++++++++++++++++++++++++++++++++++++++++++++++++++++++++++
% Abstract
\begin{abstract}
Many extensions of the Standard Model (SM) of particle
physics predict the existence of charged Higgs bosons with substantial
couplings to SM particles, which would render them observable both
directly at the LHC and indirectly at $B$-factories. For example, the
charged Higgs boson couplings to fermions in two doublet Higgs models
of type II, are proportional to the ratio of the two Higgs doublet
vacuum expectation values ($\tan\beta$) and fermionic mass factors and
could thus be substantial at large $\tan\beta$ and/or for heavy
fermions. In this work we perform a model-independent study of the
charged Higgs boson couplings at the LHC and at $B$-factories for
large values of $\tan\beta$. We have shown that at high luminosity it
is possible to measure the couplings of a charged Higgs boson to the 
third generation of quarks up to an accuracy of 10\%. 
We further argue that by combining the possible measurements of the LHC and
the $B$-factories, it is possible to perform a universality test of
charged Higgs boson couplings to quarks. 

\end{abstract}
%+++++++++++++++++++++++++++++++++++++++++++++++++++++++++++

\maketitle

\section{Introduction \label{section:1}}
In the Standard Model (SM) of particle physics the electroweak
symmetry is broken by a complex Higgs-boson doublet, resulting in one
neutral scalar particle in the physical spectrum. However, many
extensions of the SM, most notably the 2-Higgs Doublet Models (2HDM),
predict the existence of a charged Higgs boson. Recently $B$-factories
have started providing data for processes with a $\tau$-lepton in the
final state, namely $B \to D^{(*)} \tau \nu$ \cite{Aubert:2007dsa} and
$B \to \tau \nu$ \cite{Ikado:2006un}. These channels can be mediated
by a charged Higgs boson at tree-level and can provide very useful
indirect probes into the charged Higgs boson properties. Furthermore,
with the Large Hadron Collider (LHC) about to commence in earnest,
studies involving the LHC environment promise the best avenue for
directly discovering a charged Higgs boson. The issue of discovering
and measuring a charged Higgs boson's properties has been extensively
discussed in the literature (see for example Refs.
\cite{Assamagan:2002ne,Hashemi:2008ma,Mohn:2007fd,cms:tb,cms:taunu,Potter:2008zza}).
It has been shown that the $H^\pm \to \tau \nu$ mode can be used to
determine the charged Higgs boson mass at the LHC; whilst the charged
Higgs boson mass and $\tan\beta$ ($\equiv v_2/v_1$, the ratio of the
vacuum expectation values  of the two Higgs doublets) can be
determined simultaneously using the two decay modes $H^\pm \to t b$
and $H^\pm \to \tau \nu$. In this work we consider four different processes (two from the
LHC and two from $B$-factories) that could provide independent
measurements of the charged Higgs boson couplings:   
\begin{itemize}
\item{\bf LHC:} $p p \to t (b) H^+$: through the decays $H^\pm \to
\tau \nu$, $H^\pm \to t b$ ($b-t-H^\pm$ coupling)\footnote{This
extremely useful discovery channel for the charged Higgs boson allows
for a study in a wide range of $\tan\beta$ \cite{Assamagan:2002ne}.}.
\item{\bf $B$-factories: } $\btotn$ ($b-u-H^\pm$ coupling), $\btodtn$
($b-c-H^\pm$ coupling).
\end{itemize}

The processes mentioned above have several common characteristics with
regard to the charged Higgs boson couplings to the fermions. Firstly,
the parameter region of $\tan\beta$ and the charged Higgs boson mass
covered by charged Higgs boson production at the LHC ($p p \to t(b)
H^+$) overlaps with those explored with the $B$-decays at
$B$-factories. Secondly, these processes provide four independent
measurements to determine the charged Higgs boson properties. With
these four independent measurements one can in principle determine the
four parameters related to the charged Higgs boson couplings to
quarks, namely 
$\tan\beta$ and the three generic couplings related to the $b-i-H^\pm$
($i = u,c,t$) vertices. In our analysis we focus on the large
$\tan\beta$-region \cite{Hall:1993gn}, where one can neglect the term
proportional to $m_u  \cot\beta$, with $m_u$ being the up-type quark mass. At
tree-level charged Higgs boson couplings to fermions depend only on
$\tan\beta$ and the mass of the down-type fermion involved in the 
interaction. Hence, at tree-level, the $b-i-H^\pm$ ($i = u,c,t$)
vertex is the same for all the three up-type generations, that is, for
all three values of $i$. We refer to this property as the {\sl
universality} of charged Higgs boson interactions. This universality
is broken by loop corrections to the charged Higgs boson vertex.     

Our strategy is to determine the charged Higgs boson properties
first through the LHC processes. Note that the latter have been
extensively studied in many earlier works (see
Ref. \cite{Assamagan:2002ne}, for example) with the motivation of
discovering the charged Higgs boson in the region of large 
$\tan\beta$. We shall assume that the charged Higgs boson is already
observed with a certain mass. Using the two LHC processes as indicated
above, one can then determine $\tan\beta$ and the $b-t-H^\pm$
coupling. Having an estimate of $\tan\beta$ one can then study the
$B$-decays and try to determine the $b-(u/c)-H^\pm$ couplings from
$B$-factory measurements. This procedure will enable us to measure the
charged Higgs boson couplings to the bottom quark and the two other
generations of up-type quarks and hence perform a universality 
check of the couplings in a model-independent way\footnote{By
universality check we mean to check 
if all the couplings of a charged Higgs boson to the bottom quark and
three up-type quarks are identical.}.      

Our paper is organized as follows : In section \ref{sec:2} we present
the general formalism of our analysis. 
In section \ref{sec:3} we present the results of the
cross-sections at the LHC. We also give the details of the framework of
the simulations we have carried out for estimating the charged Higgs
signals at the LHC. In section \ref{sec:4} we discuss the effect of
charged Higgs on tauonic $B$-decays. In section \ref{sec:5} we have
shown various correlations between the possible observables at LHC and
$B$-factories. In addition we have given a summary of the simulation
results at the LHC that we require in measuring the charged Higgs couplings. 
Finally we conclude with our summary and conclusions in section
\ref{sec:6}. In this section we have given a particular parameter
space point to demonstrate our proposal for performing a universality
test of the charged Higgs boson couplings to quarks. 

\section{General Formalism}\label{sec:2}
Let us now briefly consider how the charged Higgs boson interacts
with fermions. Further details are given in 
Ref. \cite{Itoh:2004ye}. In 2HDMs of type II, such as the one
implemented in the minimal supersymmetric SM (MSSM) among other
possible models, also in (at least in certain limits of) those of type
III, the down-type mass matrix and the charged Higgs-boson
interactions with right-handed down-type quarks is written as:  
\begin{eqnarray}
\mathcal{L}_{\rm D-quark} = - \ODR \widehat{M}_d\DL +
\frac{\sqrt{2}}{v}\tan\beta H^- \ODR^\prime\widehat{M}_d V_{\rm
CKM}^\dag \widehat{R}_d^{-1}\UL + {\rm h.c.} \,\,\, , \label{down}   
\end{eqnarray}
where we have diagonalized the Yukawa sector of the Lagrangian and
assumed Minimal Flavour Violation (MFV) with flavour mixings given
only by the Cabibbo-Kobayashi-Maskawa (CKM) matrix. $U_{L/R}$ and
$D_{L/R}$ are the left/right-handed up- and down-type
quarks\footnote{Quark fields represented by capital letters represent
three vectors in flavour space.}, and $\widehat{M}_d$ is the diagonal
down-type mass matrix. The trilinear couplings are in general
proportional to the original Yukawa couplings, and we shall label the
components of the diagonal matrix $\widehat{R}_d^{-1} = 
\mathrm{diag} \left[R^{-1}_{11}, R^{-1}_{22} R^{-1}_{33} \right]$,
where the three diagonal values of $\widehat{R}_d^{-1}$ represent the
couplings of a charged Higgs boson to the bottom quark and the three
up-type quarks.
At tree-level, these three couplings are universal,
that is $R_{11}^{-1} = 
R_{22}^{-1} = R_{33}^{-1} = 1$. This universality is broken to some
extent by loop corrections to the charged Higgs boson vertex, and
$\widehat{R}_d$ can then be written as: 
\begin{equation}
\widehat{R}_d = 1 + \tan\beta \hat{\Delta}_{m_d} \,\,\, . 
\end{equation}

\par
\begin{figure}[htb]
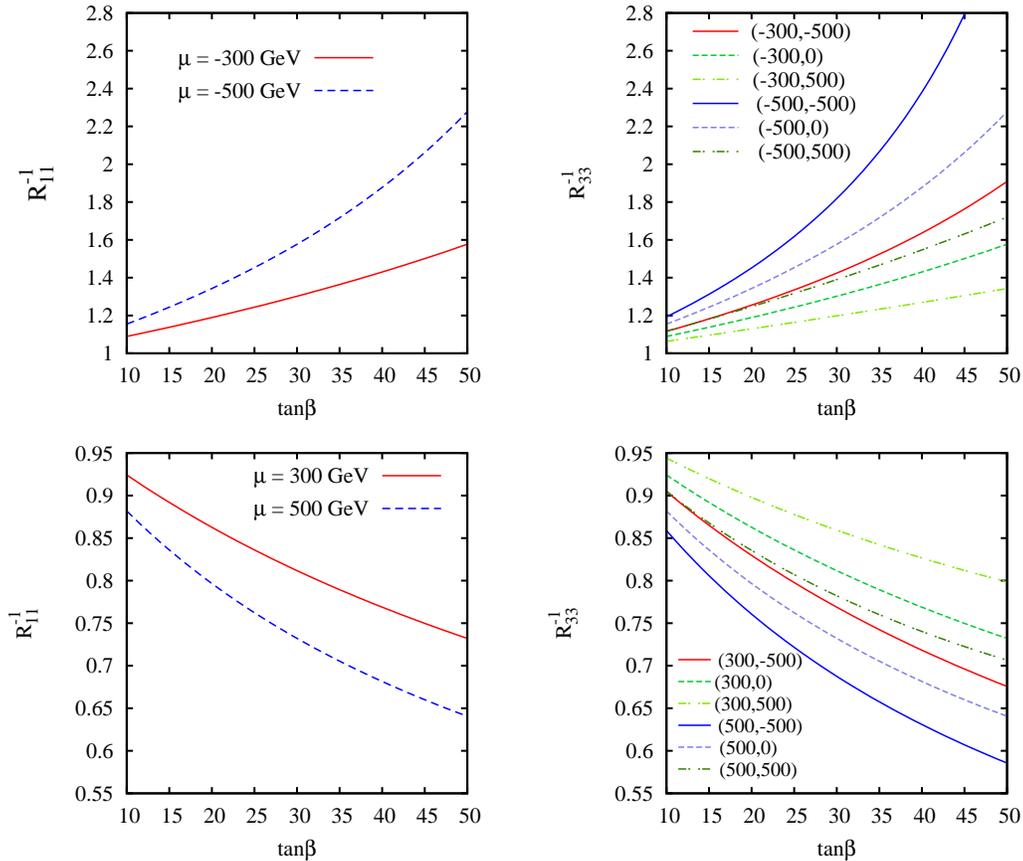

%\vskip -1cm
\bcen
\epsfig{file=R11_mu_neg,width=.55\textwidth} \hskip -2cm
\epsfig{file=R33_mu_neg,width=.55\textwidth}\\ \vskip -0.5cm
\epsfig{file=R11_mu_posi,width=.55\textwidth} \hskip -2cm
\epsfig{file=R33_mu_posi,width=.55\textwidth}
%\vskip 0.2cm
\caption{\sl Dependence of the general couplings $R^{-1}_{ii}$ on
$tan\beta$ in the exemplary case of the MSSM for various values of the
higgsino mass parameter $\mu$ and the up-type trilinear coupling
$A$. The left-hand plots are for $R^{-1}_{11} = R^{-1}_{22}$, while
those on the right are for $R_{33}^{-1}$. We 
present the case of negative $\mu$ in the top panels and for positive
$\mu$ below. The other SUSY parameters are $M_{\tilde{g}}$ = 800 GeV
and $M_{\tilde{b}_1} = M_{\tilde{t}_1} = 500$ GeV. We have 
also assumed $M_{\tilde{t}_L} = M_{\tilde{t}_R}$ and $M_{\tilde{b}_L}
=  M_{\tilde{b}_R}$. The legends in the right top and right bottom
panels correspond to $(\mu,A)$ in GeV.}  
\label{fig:RR} 
\ecen
\vskip -.7cm
\end{figure}

 In the MSSM, for example, $\hat{\Delta}_{m_d}$ includes the
contributions from the gluino and down-type squarks (SUSY-QCD) and of
the charged higgsino and up-type squarks \cite{Itoh:2004ye,Carena:1999py}.
In our analysis we have kept the ${\cal O}(\alpha_s)$ SUSY-QCD
corrections and SUSY loop corrections associated with the Higgs-top
Yukawa couplings ($y_t$) and have  
neglected the subleading electroweak corrections of the order ${\cal
O}(g^2)$ as given in Ref. \cite{Gorbahn:2009pp}.\footnote{For an alternative definition, in which SUSY
loop effects are assigned to the CKM matrix, see Ref.\ \cite{Blazek:1995nv}} The latter are calculated in the unbroken phase of the ${\rm SU}(2)\times {\rm U}(1)$
symmetry and under the assumption that soft breaking terms for squark
masses are proportional to the unit matrix in the flavour basis. Therefore, they
then depend upon the higgsino-mass parameter $\mu$, the up-type
trilinear couplings $A$, and the bino, bottom and top squark
masses. As argued in Ref.\ \cite{Itoh:2004ye} the higgsino-diagram
contributions can be neglected in $R^{-1}_{11}$ and $R^{-1}_{22}$, so
that to a very good approximation $R^{-1}_{11} \approx
R^{-1}_{22}$
\footnote{{The charged Higgs coupling in Eq.(\ref{down}) are derived 
upon the assumption that soft SUSY breaking mass matrices 
of squarks with the same gauge quantum numbers are proportional to
a unit matrix in generation space \cite{Dedes:2002er}. 
General cases of MFV are discussed in Ref. \cite{D'Ambrosio:2002ex, Buras:2002vd}
and the corresponding formula is given also in Ref. \cite{Itoh:2004ye}.
In particular, the $b-u-H^{\pm}$ and $b-c-H^{\pm}$ coupling constants
are different in general cases, but we have checked that 
the difference is numerically very small in the SUSY parameter choice 
considered in this paper.}}. 
For illustration, we show in Fig.\ \ref{fig:RR} the
dependence of the SUSY corrections on $\tan\beta$ for some
illustrative SUSY parameters. These corrections can alter the 
tree-level values significantly, although low-energy data (e.g.\ from
$b\to s\gamma$, $B - \bar{B}$ mixing, $B \to \mu \mu$ and $b\to
s\mu\mu$) restricts the  
admissible parameter space \cite{Babu:1999hn}. In addition, it can be
observed that the higgsino  
corrections are proportional to the up-type Yukawa couplings and hence
can be substantial for diagrams involving the top quark as an external
fermion line. This effectively implies that $R^{-1}_{33}$ can differ
substantially from $R^{-1}_{11}$, where for certain SUSY scenarios, as
shown in Fig.\ \ref{fig:RR}, we observe that $R^{-1}_{33}$ can differ
from $R^{-1}_{11}$ by more than 30\%. This difference could be
observed at the LHC for processes that depend on $R^{-1}_{33}$ when
compared with the results of $B$-factories for processes that depend
on $R^{-1}_{11}$. We remind the reader that the effective couplings
are invariant under a rescaling of all SUSY masses and may indeed be
the first observable SUSY effect, as long as the heavy Higgs bosons
are light enough. The situation is similar in other models predicting
a charged Higgs boson, such as those with a Peccei-Quinn symmetry,
spontaneous $CP$ violation, dynamical symmetry breaking, or those
based on $E_6$ superstring theories, but these have usually been
studied much less with respect to the constraints imposed by
low-energy data. In the remainder of this work, we shall thus treat
the diagonal entries of $\widehat{R}^{-1}_d$ as model-independent free
parameters in our simulations and numerics, but we will assume that
$R^{-1}_{11} \approx R^{-1}_{22}$.  
One can constrain these parameters from the low energy FCNC (Flavour
Changing Neutral Current) processes like $b\to s\gamma$, $B - \bar{B}$
mixing, $B \to \mu \mu$ and $b\to s\mu\mu$ in a {\sl model dependent}
framework. But, in our analysis we have considered these to be the
effective couplings of a charged Higgs boson to fermions in a model
independent framework and hence it is easy to evade the constraints
from low energy FCNC processes.  
Note that the corresponding corrections to the up-type couplings are
suppressed by $\cot\beta$ and hence can be neglected in our
analysis. The electroweak corrections to the charged leptons are given
in Ref.\ \cite{Gorbahn:2009pp} and are usually small. 

%%====================================================================
%%====================================================================

\section{Charged Higgs at the LHC}\label{sec:3}
As argued above, at the LHC we can observe the vertex $b-t-H^\pm$
by producing a charged Higgs boson. In the large $\tan\beta$ region of
the parameter space, the charged Higgs boson will be predominantly
produced through the partonic process $gb \to t H^\pm$. The effective
interaction term of the charged Higgs boson with $t$ and $b$ quarks in
the 2HDMs, that can be probed at the LHC, can be written as:   
\begin{eqnarray}
{\cal L} &= &\frac{g}{2\sqrt{2}\; m_W} V_{tb} H^+ {\bar t}
\left( m_t \cot \beta (1-\gamma_5)  + m_b R_{33}^{-1} \tan \beta
(1+\gamma_5) \right) b + h.c. \,\,\, .   
\end{eqnarray}
%%====================================================================
\begin{figure}[htb]
\vskip -.5cm
\bcen
\epsfig{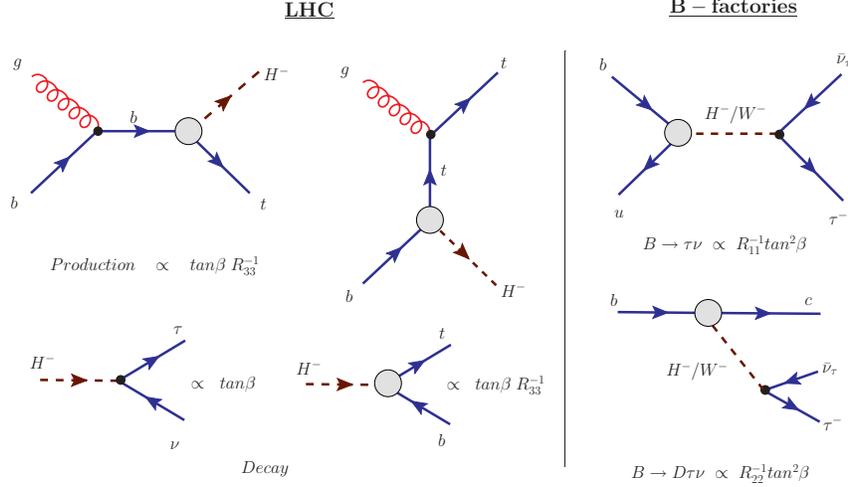}
\caption{\sl Production and decay of a charged Higgs boson at the LHC
 involving the general third-generation coupling $R_{33}^{-1}$ (left)
 and (semi-)leptonic $B$-decays involving the general first- and
 second-generation couplings $R_{11}^{-1}$ and $R_{22}^{-1}$.}
\label{fig:1}
\ecen
\vskip -.5cm
\end{figure}
%%====================================================================
The cross-section for $gb\rightarrow tH^\pm$ (see the left-hand
portion of Fig. \ref{fig:1}) can be written as:   
\begin{equation}
\label{gbcross}
\sigma(gb\rightarrow tH^\pm) \propto \left( m^2_t
\cot^2 \beta + m^2_b (R^{-1}_{33})^2 \tan^2\beta \right) \,\,\, .   
\end{equation}
In the large $\tan\beta$-region of the parameter space, the decay
width for a charged Higgs boson is dominated by the channels $H^\pm
\to \tau \nu$ and $H^\pm \to t b$, whose decay widths in the large
$\tan\beta$ region can be written as:   
\begin{eqnarray}
\Gamma (H^- \to \tau^- \nu_\tau ) &\simeq& \frac{m_{H^\pm}}{8 \pi v^2}
m_\tau^2 \tan^2 \beta \left(1 - \frac{m_\tau^2}{m_{H^\pm}^2} \right)^2
\,\,\, ,  \label{eqn:tauwidth}  \\ 
\Gamma(H^- \to {\bar t}b) & \simeq & \frac{3 \; m_{H^\pm}}{8 \; \pi
v^2}\  m_b^2 (R^{-1}_{33})^2 \tan^2 \beta 
\left( 1 -\frac{m_t^2}{m_{H^\pm}^2}-\frac{m_b^2}{m_{H^\pm}^2}\right) 
\nonumber \\   
&& \hspace{2cm} \times 
\left[ 1
-\left(\frac{m_t+m_b}{m_{H^\pm}}\right)^2 \right]^{1/2} \left[ 1
-\left(\frac{m_t-m_b}{m_{H^\pm}}\right)^2 \right]^{1/2} \,\,\, . 
\label{decayhtb}  
\end{eqnarray}

%%====================================================================
\begin{figure}[tb]
\begin{center}
\hspace*{-2cm}
\epsfig{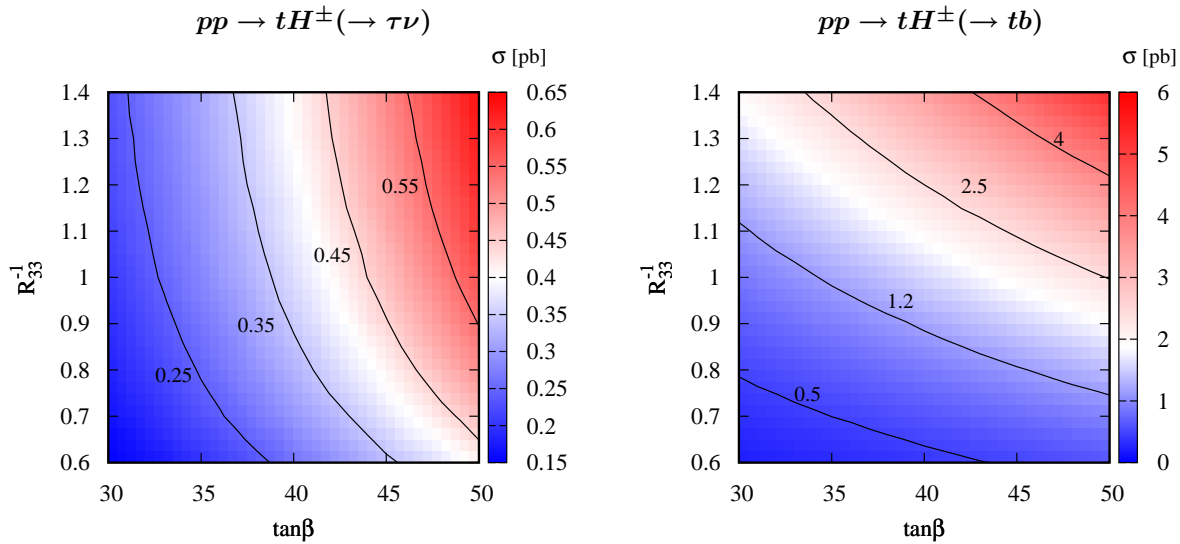}
\vskip -0.2cm
\caption{\sl Contour plots of the cross-sections for the processes $pp
\to t H^\pm (\to \tau \nu)$ (left) and $pp \to t H^\pm (\to t b)$
(right) versus $R^{-1}_{33}$ and $tan\beta$ with fixed $m_{H^\pm} =$
300 GeV.}    
\label{fig:contour}
\ecen
\vskip -.2cm
\end{figure}
%%====================================================================

For the main production mechanism ($gb\to tH^\pm$) considered
here, we simulated the leading order (LO) cross-sections using the
{\sc pythia} Monte Carlo generator \cite{Sjostrand:2006za} and {\sc 
cteq6l1} parton densities \cite{Nadolsky:2008zw} using LHAPDF
\cite{Whalley:2005nh}. The renormalisation
and factorisation scales $\mu_R$ and $\mu_F$ were identified with the
average mass of the two final state particles. It is well known that
the process $gg\to tbH^\pm$, although formally of next-to-leading
order (NLO), is numerically important due to the enhanced splitting of
gluons into collinear bottom-antibottom pairs. In our simulations
this process was taken into account 
following the ideas originally proposed in the late 80's in Ref. \cite{Barnett:1987jw},
using the {\sc matchig}
\cite{Alwall:2005gs} addition to {\sc pythia}6.4.11
\cite{Sjostrand:2006za}, which then allowed for a better description
of hard additional jets and at the same time for a consistent
subtraction of the mass logarithm coming from the overlap with the LO
process \cite{Alwall:2005gs}. However, since the matched sum is still
normalised to the LO total cross-section, we renormalise it to NLO
precision using {\sc cteq6m} parton densities and the corresponding 
value of $\lambda_{\overline{\rm MS}}^{n_f=5}= 226$ MeV in the
computations given in Ref.\ \cite{Plehn:2002vy,Berger:2003sm}, which
has been shown to be in good agreement with the one performed in Ref.\
\cite{Zhu:2001nt}. For a Higgs boson mass of 300 GeV and in the
$\tan\beta$ region of 30--50 considered here, the correction varies
very little and can be well approximated with a constant factor of 
$1.2$.     

In Fig.\ \ref{fig:contour} we show contour plots of the
cross-sections at the LHC in the $R_{33}^{-1} - \tan\beta$ plane. It
is evident from these plots and Eqs.\ (\ref{gbcross}),
(\ref{eqn:tauwidth}) and (\ref{decayhtb}) that a measurement of the
ratio of these cross-sections could give us a measurement of  
$R_{33}^{-1}$ and thus constrain the parameter space of any new
particles contributing to its loop corrections. The ratio will also be
relatively free from the theoretical uncertainty arising at the
production level of the charged Higgs boson. In principle, by
measuring these cross-sections simultaneously, one can estimate
$R_{33}^{-1}$ and $\tan\beta$. We shall demonstrate this in an
upcoming work using detailed simulations \cite{future}.    

\section{Charged Higgs and tauonic $B$-decays}\label{sec:4}
The first measurement of tauonic $B$-decays was done by the L3 collaboration
at LEP \cite{Acciarri:1994hb}, where they measured the inclusive branching
fraction of tauonic $B$-decays, namely $B \to X \tau \nu$. Based on the
LEP measurement of inclusive tauonic decays, Grossman {\sl et al.}
\cite{Grossman:1994ax} obtained the limits on $\tan\beta$ as
$\tan\beta<0.52(m_H/1 GeV)$. 
The tauonic $B$-decays can be mediated by tree level charged Higgs
boson exchanges and could be very sensitive to charged Higgs boson
couplings
\cite{Itoh:2004ye,Kiers:1997zt,Nierste:2008qe,Hou:1992sy,Tanaka:1994ay}.   
In Fig.\ \ref{fig:1} we have shown the Feynman diagrams contributing
to these decays. The importance of purely tauonic $B$-decays ($\btotn$) in constraining
charged Higgs properties was pointed out in Ref. \cite{Hou:1992sy}. The
exclusive tauonic decay $\btodtn$ offers many more kinematical
distributions \cite{Tanaka:1994ay} that could further help in pinning
down the structure of the effective Hamiltonian responsible for these decays. 
In Ref. \cite{Itoh:2004ye} it was shown that considering both these decay modes, namely
$\btotn$ and $\btodtn$, could possibly give a two fold ambiguity in
measuring $\tan\beta/m_H$. But by combining these measurements one can
possibly remove this two fold ambiguity. Using the effective vertices given in 
Eq.\ (\ref{down}) and shown in Fig.\ \ref{fig:1} (right), the rates of
the $\btotn$ and $\btodtn$ processes can be written as functions of
the parameters $\tan\beta$ and $m_{H^\pm}$ and the loop
correction factors as: 
\bea
Br(\btotn) & \propto &  \left[a + b \frac{\tan^2\beta
(R^{-1}_{11})^2}{m_{H^\pm}^2} \right]^2 \label{eq:1} \,\,\, , \label{br:1} \\   
Br(\btodtn) & \propto &  a' + b' \frac{\tan^2\beta
(R^{-1}_{22})^2}{m_{H^\pm}^2}  + c' \left(\frac{\tan^2\beta 
(R^{-1}_{22})^2}{m_{H^\pm}^2} \right)^2 \label{eq:2} \,\,\, ,   
\label{br:2}
\eea
where $a,b, a', b'$ and $c'$ are factors that do not depend on any new
physics parameters. 
In writing Eqs. (\ref{br:1}) and (\ref{br:2}) we have used the matrix
elements as given in Refs. \cite{Itoh:2004ye,
Gorbahn:2009pp,Kiers:1997zt}. The coefficients $ a', b', c'$
depend on the form factors for the $B \to D$ transition. Presently these
form factors have substantial theoretical uncertainties which can be
translated into theoretical errors on the branching ratios of
these modes. For our illustrative purposes we have used the
prescription of the form factors for $B \to D$ transition as given in Ref. 
\cite{Gorbahn:2009pp}. Gorbahn {\sl et al.} \cite{Gorbahn:2009pp} have
also estimated the possible theoretical uncertainties on the
form-factors for $B \to D$ transitions.  

\section{Effective couplings determination}\label{sec:5}
We have discussed in sections \ref{sec:3} and \ref{sec:4} the phenomenology of a charged Higgs 
boson at the LHC and at $B$-factories for large values of $\tan\beta$ in relation 
to the study of its couplings. In the following we give the results of our numerical study 
and in particular of the possibility of combining these two kinds of complementary experimental 
results in order to constrain the charged Higgs coupling.
 
%%====================================================================
\begin{figure}[tb]
\begin{center}
\hspace*{-2cm}
\epsfig{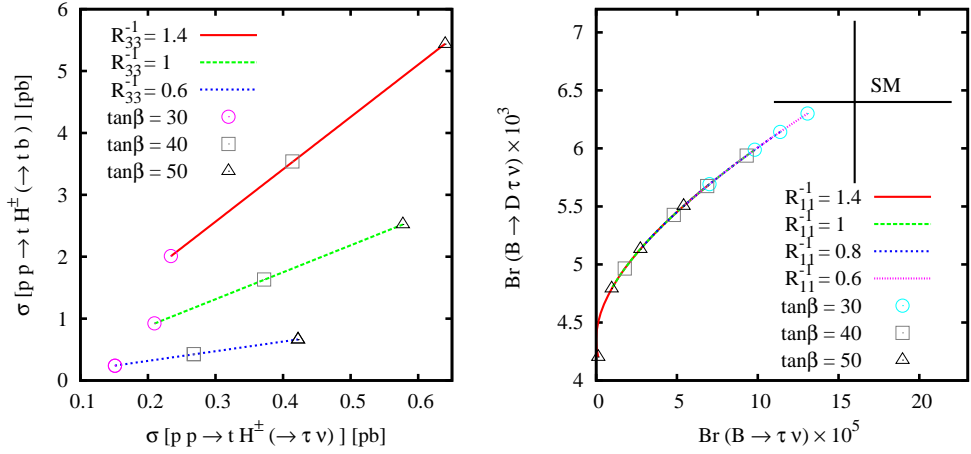}
\vskip -0.2cm
\caption{\sl Correlation plots of the cross-sections for the processes
$pp \to t (b) H^\pm (\to \tau \nu)$ and $pp  \to t (b) H^\pm (\to t
b)$ for three values of $R^{-1}_{33}$ and $tan\beta$ (left) and of the
branching ratios for $B \to D \tau \nu$ and $B \to \tau \nu$ (right)
for various values of $tan\beta$ and $\widehat{R}_d^{-1}$ with fixed
$m_{H^\pm}= 300$ GeV.}     
\label{fig:LHC-corr} 
\ecen
\vskip -.5cm
\end{figure}
%%====================================================================

%%====================================================================
\begin{figure}[tb]
\bcen
\epsfig{file=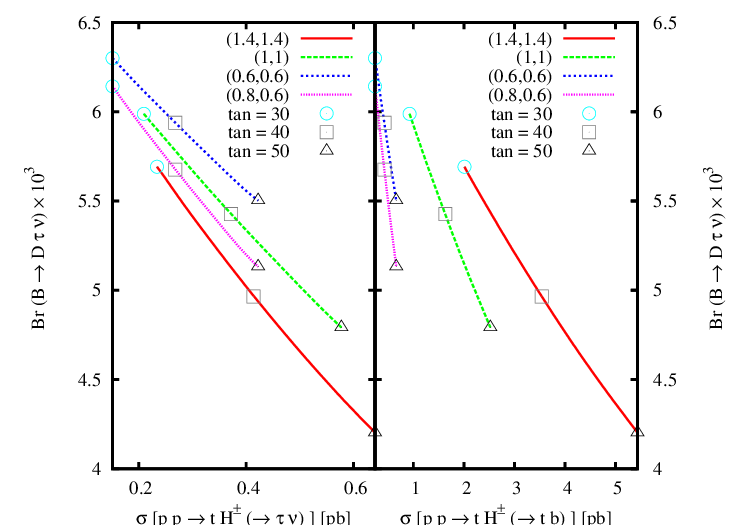,width=.5\textwidth}  %\hskip -1cm
\epsfig{file=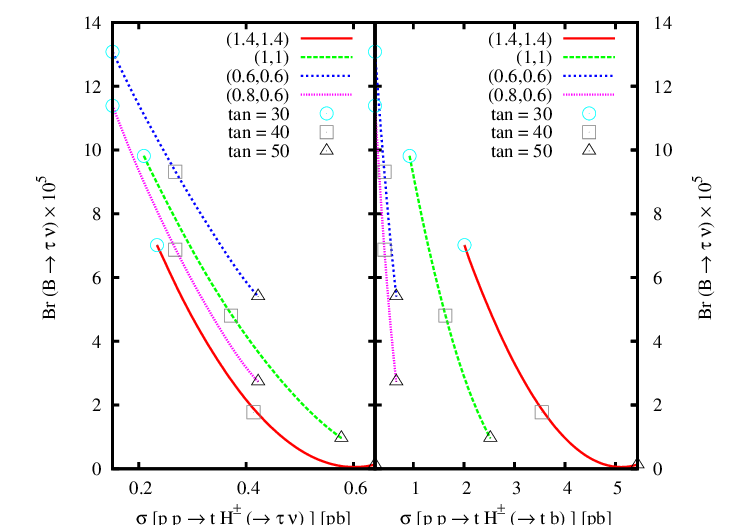,width=.5\textwidth}  
\caption{\sl Correlation plots of the branching ratio $B \to D \tau
\nu$ with the cross-sections $\sigma (pp \to t H^\pm (\to \tau \nu))$
(top left) and $\sigma (pp \to t H^\pm (\to t b))$ (top right) and of
the branching ratio $B \to \tau \nu$ with the cross-sections $\sigma
(pp \to t H^\pm (\to \tau \nu))$ (bottom left) and $\sigma (pp \to t
H^\pm (\to t b))$ (bottom right) for various values of $tan\beta$ and
$\widehat{R}_d^{-1}$.  
The numbers in the legend refer to $(R_{11}^{-1},R_{33}^{-1})$.} 
\label{fig:Combined-corr}   
\ecen
%\vskip -.5cm
\end{figure}
%%====================================================================

In Figs. \ref{fig:LHC-corr} and \ref{fig:Combined-corr} we have
plotted the correlations amongst the four processes (two from the LHC
and two from $B$-factories) considered here. In these plots we have
varied $\tan\beta$ in the range $30 < \tan\beta < 50$ for different
values of $R_{ii}^{-1}$ ($ii = 11, 33$). The left panel of
Fig. \ref{fig:LHC-corr} shows the correlation of the LHC observables,
whilst the correlation of $B$-decay branching ratios in the right
panel of Fig. \ref{fig:LHC-corr} gives the same line for different
values of $R_{11}^{-1}$. The reason for this can be seen from the
Feynman diagrams given in Fig. \ref{fig:1} where $R_{ii}^{-1}$ and
$\tan\beta$ arise from the same combination ($\equiv R_{ii}^{-1}
\tan^2\beta$) in the tauonic $B$-decays considered in this work. Hence
the measurement of these two $B$-decays will only give an estimate of
the product of $R_{11}^{-1}$ and $\tan\beta$. However, by considering
the correlations of the $B$-decay observables with LHC observables, as
shown in Fig. \ref{fig:Combined-corr}, one can remove this
degeneracy. So in principle it is possible to measure the four
parameters ($\tan\beta$ and $R_{ii}^{-1}$ with $ii=11, 22, 33$) using
the six correlation plots shown in Figs. \ref{fig:LHC-corr} and
\ref{fig:Combined-corr}. We will now demonstrate this statement by
considering an illustrative set of input parameters.   

The illustrative point we have chosen lies at $\tan\beta = 40$,
$m_{H^\pm} = 300$ GeV and $R_{33}^{-1} = 1$. For this illustrative
point we first start by summarising our simulation results for the
charged Higgs boson production and decay in the $H^\pm \to \tau \nu$
mode. As mentioned earlier, this simulation was carried out using {\sc
matchig} \cite{Alwall:2005gs} so as to include all the relevant
processes in {\sc pythia} 6.4.11 \cite{Sjostrand:2006za}. In these
simulations the charged Higgs boson is allowed to decay via the
channel $H^\pm \to \tau^{had} \nu$.\footnote{$\tau^{had}$ means the
$\tau$ is allowed to  decay only in hadronic decay modes, for which we
have used $BR( \tau \to {\rm hadrons}) = 0.65$.} The events thus
generated were further passed through the fast detector simulator {\sc atlfast}
\cite{atlfast}. {\sc atlfast} identifies isolated leptons, $b$ and
$\tau$ jets. It also reconstructs missing energy. The jets in {\sc
  atlfast} are reconstructed using simple cone algorithms. The decay
chain we have used is \cite{Assamagan:2002ne,Hashemi:2008ma,cms:taunu,Mohn:2007fd}: 
$$
p p \to t (\to j j b) (b) H (\to \tau \nu) \to j j b (b) \tau \et
$$
 For the analysis we have used following pre-selection cuts:
\begin{itemize}
\item{} The event should have exactly one $\tau$-jet with $p_T^\tau >
  40$ GeV.
\item{} There should be at least three light jets with $p_T > 30$
  GeV. At least one of these jets must be a $b$-jet. 
\end{itemize}
In order to reduce the backgrounds we have in addition used the following
selection cuts \cite{Assamagan:2002ne,Hashemi:2008ma,cms:taunu,Mohn:2007fd}:
\begin{itemize}
\item{} ${\cal N}_1$ : $p_T^\tau > 100$ GeV. This reduces the backgrounds where
  $\tau$ comes from a decay of the $W$-boson. This cut is too severe for a
  light charged Higgs but there is no substantial reduction in event
  rates for the signal as we have considered a relatively heavy Higgs. 
\item{} ${\cal N}_2$ : $\et > 100$ GeV. This cut again reduces SM backgrounds with
  neutrinos in the final state.
\item{} ${\cal N}_3$ :$\triangle \phi > 1$ radian, where $\triangle
\phi$ is the azimuthal angle between the $\tau$-jet and $\not\!\! p_T$. 
\end{itemize}
The uncertainty in cross-section measurements is estimated as \cite{Assamagan:2002ne}:
$$
\frac{\triangle (\sigma \times BR)}{ (\sigma \times BR)} =
\sqrt{\frac{S + B}{S^2}} 
$$
where $S$ and $B$ are signal and background events respectively. 

The numerical results of our analysis are summarized in
Table \ref{table:1}. The table shows that for a reasonable range of
input parameters the cross-sections at the LHC can be measured with a
10\% accuracy for a luminosity of ${\cal L}$ = 100 fb$^{-1}$, whereas
the measurement can be improved substantially for higher
luminosities. Note that the error in the measurement of $\tan\beta$ is
consistent\footnote{These numbers can change slightly if we consider
the systematic error arising from the luminosity uncertainty.} with
the observations made in Ref. \cite{Assamagan:2002ne}. For our
analysis we have taken the error in the measurement of the
cross-section in this channel to be 10\% for a luminosity of 100
fb$^{-1}$ and 7.5\% for a luminosity of 300 fb$^{-1}$.  
At this point we would like to note that for our results we have used
fast detector simulator {\sc atlfast} \cite{atlfast} and have followed
the methodology as given in Ref. \cite{Assamagan:2002ne}. In our work instead of
using {\sc pythia} to consider the dominant partonic process $gb \to t
H^\pm$ (as was done in Ref. \cite{Assamagan:2002ne})  we have used {\sc
matchig} \cite{Alwall:2005gs} in addition to {\sc pythia} to  
include the additional sub-processes namely $gg \to t b H^\pm$ and
$q\bar{q} \to t b H^\pm$. In a realistic simulation with full detector
simulator the errors in cross-section measurements could be a bit
larger. As mentioned, our results are consistent with the earlier
studies \cite{Assamagan:2002ne} and the recent studies done in Refs. \cite{cms:taunu,Mohn:2007fd} with fast detector simulators. Mohn {\sl et al.}
\cite{Mohn:2007fd} also proposed some additional set of cuts like the
transverse momentum ratio of $\tau$-jet and hardest parton jet that
has been used for top-reconstruction. This further improves the
discovery reach of charged Higgs boson in $H^\pm \to \tau \nu$ mode.  
But these studies assumed tree level {\sl Universal} coupling of
charged Higgs bosons to quarks. 

%  Table
\begin{table}[htb]
\bcen
\vskip -.2cm
\caption{\sl Cumulative Efficiencies of cuts and estimated errors for
measurements of a signal cross-section for the process
$p p \to t (b) H (\to \tau^{had} \nu)$. For these numbers we have fixed $m_{H^\pm} = 300$  
GeV.} \vskip .2cm 
\begin{tabular}{c || c | c | c} \hline
  & $R_{33}^{-1}$ = 0.7 & $R_{33}^{-1}$ = 1 & $R_{33}^{-1}$ = 1.3 \\
\hline \hline
$\sigma$ (fb)     &  204      & 249     & 273     \\
Pre-selection & 48 $\times 10^{-3}$ & 48 $\times 10^{-3}$ & 48 $\times 10^{-3}$ \\ 
${\cal N}_1$ & $12.8 \times 10^{-3}$ & $13 \times 10^{-3}$ & $13 \times 10^{-3}$ \\
${\cal N}_2$ & $61 \times 10^{-4}$ & $67 \times 10^{-4}$ & $66 \times 10^{-4}$ \\ 
${\cal N}_3$ & $47 \times 10^{-4}$  &  $53\times 10^{-4}$ &  $52 \times 10^{-4}$ \\
$\triangle \sigbr/\sigbr$ (${\cal L} = 100 fb^{-1}$) & 10.6 $\%$ & 9.5
$\%$ & 8.6 $\%$ \\   
$\triangle \sigbr/\sigbr$  (${\cal L} = 300 fb^{-1}$) & 6.2 $\%$ & 5.5
$\%$ & 5 $\%$ \\  
\hline
\end{tabular}
\label{table:1}
\ecen
\vskip -.4cm
\end{table}
%%====================================================================
\begin{figure}[htb]
\vskip 1cm
\bcen
\hskip -1cm
\epsfig{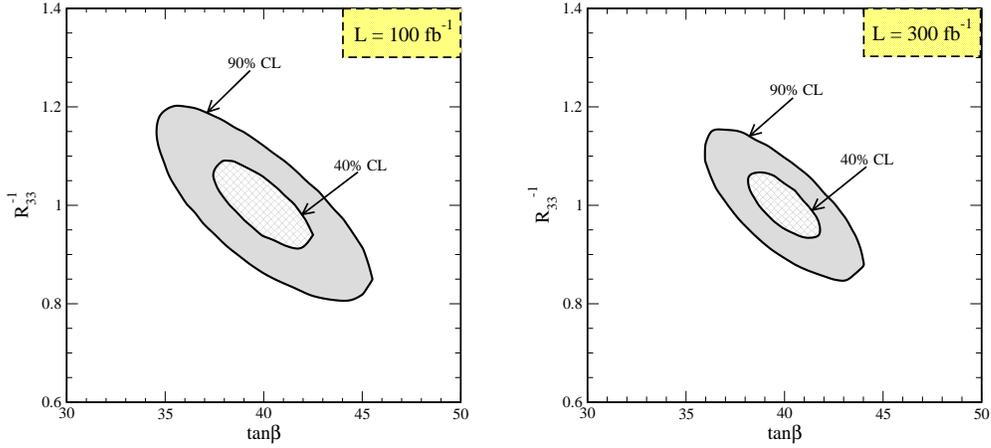}
\vskip .4cm
\caption{\sl $\chi^2$ contours for 100 fb$^{-1}$ (left panel) and 300
fb$^{-1}$ (right panel) luminosities with input values $R_{33}^{-1}=1$ and $\tan \beta =40$.}  
\label{chisq-cont}
\ecen
\vskip -.5cm
\end{figure}
%%====================================================================

%%====================================================================
\begin{figure}[htb]
\bcen
\hskip -1cm
\epsfig{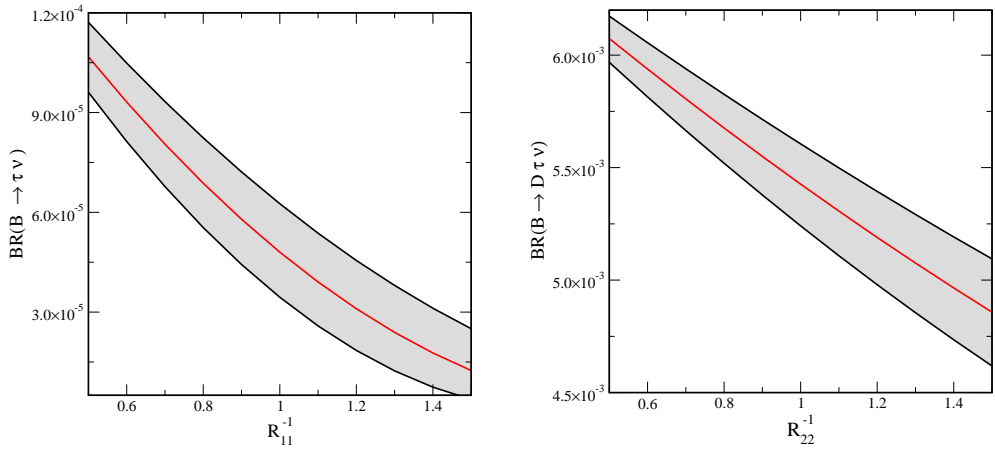}
\vskip .2cm
\caption{\sl Branching ratios of $B \to \tau \nu$ (left) and $B \to D
\tau \nu$ (right) as a function of $R_{11}^{-1}$ and $R_{22}^{-1}$,
respectively. The shaded area corresponds to $tan\beta$ in the range
$37 
< tan\beta < 43$.}  
\label{bdecay}
\ecen
\vskip -.5cm
\end{figure}
%%====================================================================

For the other dominant decay mode of the charged Higgs boson,
namely $H^\pm \to t b$, the cross-section times branching ratio is
much larger than for $H^\pm \to \tau \nu$. However, this mode has at
least three $b$-jets in its final state. To detect a charged Higgs
boson in this mode one has to tag all the final state $b$-jets, which
reduces the efficiency of this mode. In this mode one therefore has to
reconstruct the two top-quark candidates in order to reconstruct a
charged Higgs boson by using the invariant mass $m_{tb}$. The
combinatorial backgrounds associated with $t \bar{t}$ production makes
the observation of this channel a challenging task
\cite{Assamagan:2002ne}. For this reason the $H^\pm \to t b$ channel
may not be a good discovery channel for a charged Higgs boson at the
LHC. ATLAS \cite{Assamagan:2002ne,Potter:2008zza} and CMS
\cite{cms:tb} studies for a discovery of the charged Higgs boson using
this decay mode indicate that, although it is possible to observe the
charged Higgs boson in this mode, the most optimistic scenario for the
discovery of a charged Higgs boson at the LHC in the large $\tan\beta$
parameter space is through the decay $H^\pm \to \tau \nu$
\cite{Hashemi:2008ma,Mohn:2007fd,cms:taunu,Potter:2008zza,Assamagan:2002ne}.
However, as argued in Ref. \cite{Assamagan:2002ne}, the significance
of this channel improves if one has a prior estimation of the mass of
the charged Higgs boson. It was emphasised in
Ref. \cite{Assamagan:2002ne} that the channel ($H^\pm \to t b$) could
be a useful channel to probe the charged Higgs boson for a low charged
Higgs boson mass (up to a charged Higgs boson mass of 300 GeV) in a
relatively low $\tan\beta$ region ($\tan\beta < 2.5$). For high
$\tan\beta$ ($> 25$), the charged Higgs boson can be probed in this
channel for a fairly high charged Higgs-boson mass. The detection of a
heavy charged Higgs boson in this mode is possible because we can have
a hard $b$-jet in the final state that can help in optimising the
cuts, and hence improving the efficiency. For our analysis we have
taken the accuracy of the cross-section in this 
channel, that is, the error in the measurement of the $p p \to t H^\pm
(\to t b)$ cross-section to be 15\% for a luminosity of 100 fb$^{-1}$
and 12\% for a luminosity of 300 fb$^{-1}$.    

\section{Discussion and conclusions} \label{sec:6}
By measuring the cross-sections of the two decay modes of the
charged Higgs boson at the LHC we discussed in the previous section, 
one can determine the charged Higgs
boson couplings. Assuming the charged Higgs boson mass to be known
(taken to be 300 GeV in our present analysis) we have tried to do a
$\chi^2$ fit in the $R_{33}^{-1}$ - $\tan\beta$ plane using the
cross-section measurement uncertainties as given in Table \ref{table:1} 
and at the end of the previous section. The results
are presented in Fig. \ref{chisq-cont}. As can be seen from these
figures, it might be possible to measure $R^{-1}_{33}$ and $\tan\beta$
with an accuracy of about 10\% at high luminosity. Armed with this
information about $\tan\beta$, from the LHC measurements, it can then
be taken as an input to the $B$-decay measurements, namely $\btotn$
and $\btodtn$. In Fig. \ref{bdecay} we have plotted the respective
branching ratios of the $B$-decays as a function of $R_{11}^{-1}$ and
$R_{22}^{-1}$. The shaded area in this figure corresponds to a 10\%
error in $\tan\beta$\footnote{In Ref. \cite{Assamagan:2002ne} it was
inferred that for large values of $\tan\beta$ ($\geq 40$),
measurements to a precision of 6-7\% for high luminosity LHC results
are possible. Our results are consistent with these observations.}
around the central value of 40. Future Super-$B$ factories are
expected to measure the $\btotn$ and $\btodtn$ to a precision of 4\%
and 2.5\% respectively \cite{Browder:2007gg}. The present world
average experimental results for tauonic $B$-decays are $BR(B \to \tau
\nu) = (1.51 \pm 0.33) \times 10^{-4}$ and $BR(B \to D \tau \nu)/BR(B
\to D \mu \nu) = (41.6 \pm 11.7 \pm 5.2) \% $
\cite{Aubert:2007dsa,Browder:2007gg}. 
Presently if one uses UTfit prescription of $|V_{ub}|$ then there is substantial disagreement between experimental and SM estimates for the
branching fractions of $\btotn$. However, the PDG value of $|V_{ub}|$ is
much higher than the value given by UTfit. The main source of this
problem is the large difference in the inclusive and exclusive
estimation of $|V_{ub}|$ \cite{Lunghi:2009ke}.  
Recently, proposals have been given in Ref. \cite{Lunghi:2009ke} to reduce
this tension between experimental and theoretical SM values of
$\btotn$. To sum up the present theoretical
uncertainties in $BR(B \to \tau \nu)$ and $BR(B \to D \tau \nu)$ 
\footnote{Recently Nierste {\sl et al.} \cite{Nierste:2008qe}
attempted to update the form factors for $\btodtn$. This update
reduced the theoretical errors on vector form factors
substantially. The results we have presented use the central values
of the form factors as given in Ref. \cite{Nierste:2008qe}.} 
there is still a large discrepancy due to $f_B$, $|V_{ub}|$ and semi-leptonic form factors,
where it may be possible to reduce these uncertainties to a 5\% level
in the future \cite{bfuture}. Transforming the improved projected
theoretical information of these decays along with future Super-$B$
factory measurements in Fig. \ref{bdecay} one can measure
$R_{11}^{-1}$ and $R_{22}^{-1}$ to a fairly high precision.    

To summarise, we have tried to demonstrate that at the LHC alone
it is possible to measure the charged Higgs boson couplings, namely
$\tan\beta$ and $R^{-1}_{33}$, to an accuracy of less than
10\%. Combining this information from the LHC with improved
$B$-factory measurements, one can measure all four observables
indicated in the introduction. These observables represent effective
couplings of a charged Higgs boson to the bottom quark and the three
generations of up-type quarks, thus demonstrating that it is possible
to perform a universality test of the charged Higgs boson couplings
to quarks by the combination of low energy measurements at future
Super-$B$ factories and charged Higgs boson production at the LHC.

%%%%%%%%%%%%%%%%%%%%%%%%%%%%%%%%%
%  Acknowledgements

\section*{Acknowledgments}\label{sec:ackn}\cleqn
We would like to thank J. Alwall, T. Iijima, S. Nishida, Z. Was,
M. Worek and Andrew Akeroyd for useful comments and 
discussions. The research of A.D. and M.K. are supported in part by
the ANR project SUSYPHENO 
(ANR-06-JCJC-0038). The research of Y.O. is supported in part by the
Grant-in-Aid for Science Research, Ministry of Education, Culture,
Sports, Science and Technology (MEXT), Japan, No. 16081211 and by the
Grant-in-Aid for Science Research, Japan Society for the Promotion of
Science (JSPS), No. 20244037. 
%% NG
NG acknowledges support in part from Japan Society for Promotion of
Science (JSPS) and University Grants Commission 
(UGC) India under project number 38-58/2009 (SR).
This collaboration was made possible by
funding by the French ``Minist\`ere des Affaires Etrang\`eres'' and
the Japanese JSPS under a PHC-SAKURA project.  

%%%%%%%%%%%%%%%%%%%%%%%%%%%%%%%%%
% Bibliography

\end{document}